\documentclass[conference]{IEEEtran}
\IEEEoverridecommandlockouts

\usepackage{cite}
\usepackage{amsmath,amssymb,amsfonts}
\usepackage{algorithmic}
\usepackage[linesnumbered,ruled]{algorithm2e}
\usepackage{amsthm}
\usepackage{array}
\usepackage{graphicx}
\usepackage{textcomp}
\usepackage{paralist}
\usepackage{subfigure}
\usepackage{hyperref}
\usepackage{multirow}
\usepackage{url}
\usepackage{setspace}
\usepackage{tabu}
\usepackage{tablefootnote}
\usepackage{romannum}
\usepackage{enumitem}
\usepackage{arydshln}
\usepackage[normalem]{ulem}
\usepackage{threeparttable}
\usepackage{url}
\usepackage{xcolor}
\usepackage{xspace}
\usepackage{soul}
\usepackage{booktabs}
\usepackage{tabularx}
\usepackage[most]{tcolorbox}
\usepackage{multirow}

\definecolor{background}{HTML}{FEFBE7}
\definecolor{edge}{HTML}{DEA057}
\newtcolorbox{mybox}{colback=background!35,
colframe=edge,
width=\columnwidth,
arc=2mm,
auto
outer
arc
}

\setlist{leftmargin=18pt,topsep=2pt}

\newcommand{\name}{DMS\xspace}
\newcommand{\aid}{AID\xspace}
\newcommand{\company}{Huawei Cloud\xspace}

\begin{document}

\title{Managing Service Dependency for Cloud Reliability: The Industrial Practice \thanks{This work was supported by Key-Area Research and Development Program of Guangdong Province (No. 2020B010165002), Key Program of Fundamental Research from Shenzhen Science and Technology Innovation Commission (No. JCYJ20200109113403826), and the Research Grants Council of the Hong Kong Special Administrative Region, China (CUHK 14210920). }}

\author{
  \IEEEauthorblockN{
    Tianyi Yang\IEEEauthorrefmark{1},
    Baitong Li\IEEEauthorrefmark{1},
    Jiacheng Shen\IEEEauthorrefmark{1},
    Yuxin Su\IEEEauthorrefmark{2},
    Yongqiang Yang\IEEEauthorrefmark{3}, and
    Michael R. Lyu\IEEEauthorrefmark{1}
  }

  \IEEEauthorblockA{\IEEEauthorrefmark{1}Department of Computer Science and Engineering, The Chinese University of Hong Kong, Hong Kong SAR. \\ Email: \{tyyang, btli, jcshen, lyu\}@cse.cuhk.edu.hk}

  \IEEEauthorblockA{\IEEEauthorrefmark{2}School of Software Engineering, Sun Yat-Sen Univeristy, Zhuhai, China. Email: suyx35@mail.sysu.edu.cn}

  \IEEEauthorblockA{\IEEEauthorrefmark{3}Computing and Networking Innovation Lab, Huawei Cloud, Shenzhen, China. Email: yangyongqiang@huawei.com}
}


\maketitle


\begin{abstract}

    Interactions between cloud services result in service dependencies.
    Evaluating and managing the cascading impacts caused by service dependencies is critical to the reliability of cloud systems.
    This paper summarizes the dependency types in cloud systems and demonstrates the design of the Dependency Management System (\name), a platform for managing the service dependencies in the production cloud system.
    \name features the full-lifecycle support for service reliability (i.e., initial service deployment, service upgrade, proactive architectural optimization, and reactive failure mitigation) and refined characterization of the intensity of dependencies.

\end{abstract}


\section{Background and Motivation}
\label{sec:back_moti}

Modern cloud systems, including Huawei Cloud, are often constructed from a complex and large-scale hierarchy of distributed software modules.
The common practice is to develop and deploy these software modules as \emph{cloud microservices} that collectively comprise multiple \emph{cloud services}~\cite{aws-architecture}, e.g., resource allocation, virtual network management, and virtual machine management.
Different microservices serve different functionalities.
The microservices communicate through well-defined APIs and respond to external requests as a whole through service invocations.

Such an architecture benefits scalability, robustness, and agility but also complicates system reliability engineering.
However, the interactions between services cause dependencies, resulting in the cascading impact on the system.
Despite various fault-tolerance mechanisms introduced, it is still possible for minor anomalies to magnify their impacts and escalate into system outages.
When a cloud service or microservice enters an anomalous status, the anomaly can cascadingly propagate through the service-calling structure, causing a degraded user experience or even a service outage~\cite{aid}.

The cascading impacts hinder system operation and maintenance, deteriorating customer satisfaction.
For instance, during the initial service deployment or service upgrade, all the services it relies on should be ready.
During the failure mitigation and recovery, the cascading impact will slow the recovery.
Therefore, evaluating and managing the cascading impacts caused by service dependencies is crucial.

\begin{figure}[t]
    \centering
    \includegraphics[width=1\columnwidth]{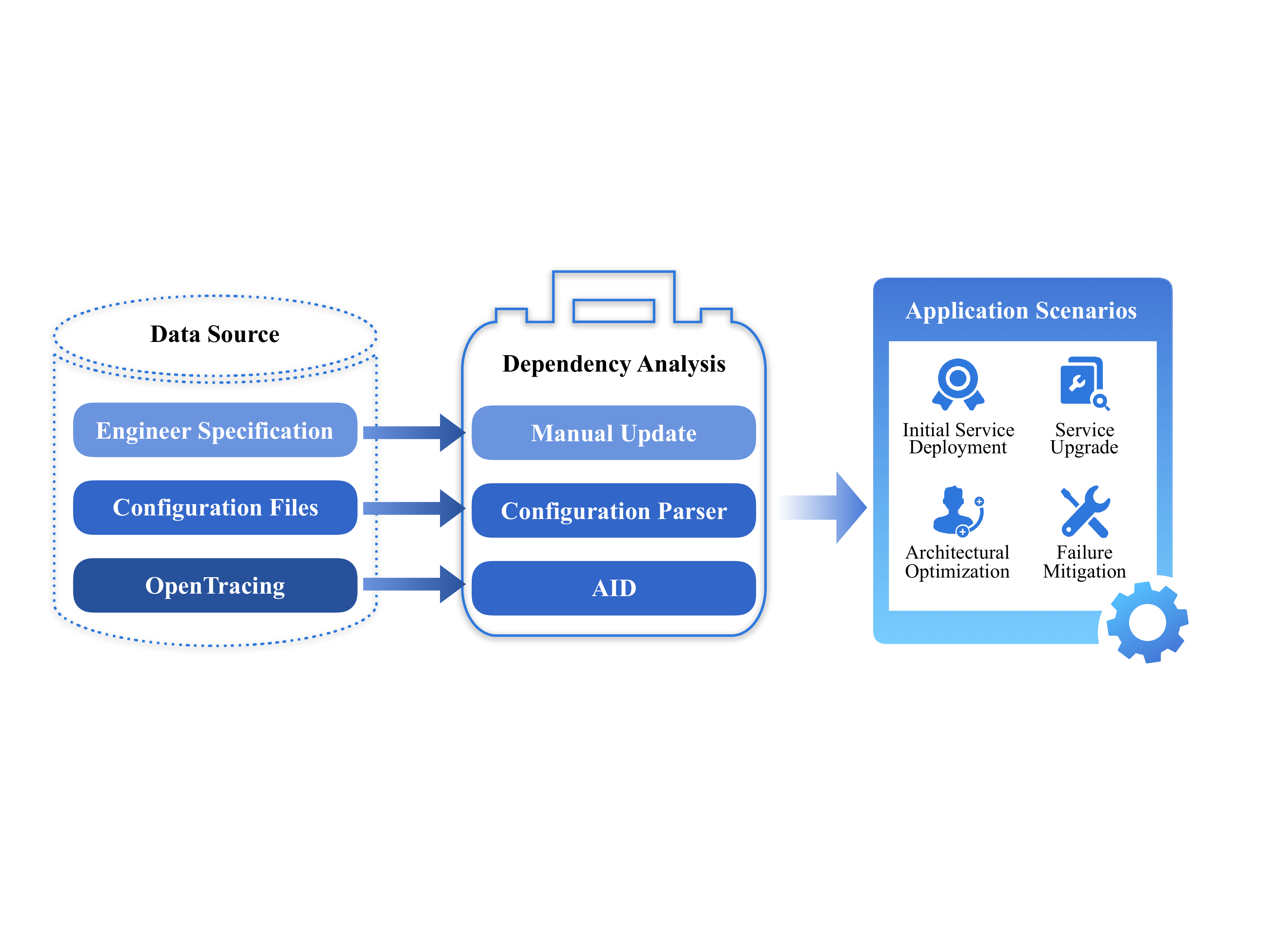}
    \vspace{-2em}
    \caption{The architecture of \name.}
    \vspace{-1.5em}
    \label{fig:arch}
\end{figure}

\section{Key Innovations}
\label{sec:innovation}

This paper classifies the dependency types in cloud systems and demonstrates the design of the Dependency Management System (\name), an end-to-end platform for managing the service dependencies in the production cloud system.
\name supports the full-lifecycle support for service reliability, i.e., initial service deployment, service upgrade, proactive architectural optimization, and reactive failure mitigation.
\name integrates our previous study on the aggregated intensity of service dependency~\cite{aid} to characterize the degree of cascading impacts and provides a refined characterization of dependencies.
In addition, \name also features automatic configuration parsing and multi-source dependency fusion for practicality.

\section{Dependency Types}
\label{sec:types}

The dependency relations in a cloud system are diverse.
In \company, we categorize the dependencies according to the architectural level, i.e., service-level dependencies and microservice-level dependencies.

\subsubsection{Service-level dependency}
If the dependency is between two cloud services, we call it a service-level dependency.
Service-level dependency can be further divided into the following three subtypes, i.e., deployment dependency, run-time dependency, and operational dependency.

\textbf{\emph{Deployment dependency}} indicates dependency during the deployment of a cloud service.
The deployment phase may rely on some cloud services to create and configure resources.
For example, the elastic computing service depends on the API management service to register public APIs.
The elastic computing service also depends on the block storage service to allocate the required resource.

\textbf{\emph{Service run-time dependency}} indicates the dependency required for the cloud service to run normally.
When a cloud service is running, it may rely on other cloud services.
For instance, the relational database service runs on the virtual machines created and managed by the elastic computing service.
Many services (e.g., the Kubernetes service) require the API management service to expose APIs to customers.
The distinction between deployment dependency and run-time dependency lies in the timing of the impact.
If the failure of the dependent service only impacts the deployment phase, the dependency is a deployment dependency.
If the impact of the failure affects the run-time functionality, the dependency is a run-time dependency.

\textbf{\emph{Operational dependency}} is the dependency required by the manual and automatic operations.
For example, the elastic computing service relies on the cloud monitoring service to monitor the entire cloud system.
This subtype of dependency usually indicates weak cascading impacts because the core functionalities will not be affected.

\subsubsection{Microservice-level dependency}
Apart from the dependency between cloud services, the microservices of one cloud service also closely interact with each other, which causes microservice-level dependency relations.
We divide the microservice-level dependency into composed-of dependency, run-on dependency, and microservice run-time dependency.
The dependency relation between a cloud service and the microservices that comprise it is the \textbf{\emph{composed-of dependency}}.
The composed-of-dependency indicates the static architecture of the cloud system.
The dependency relation between a microservice's instance and the underlying run-time environments (e.g., virtual machine, container) is the \textbf{\emph{run-on dependency}}.
The run-on dependency reflects the run-time architecture of the cloud system.
Lastly, similar to the service run-time dependency, the \textbf{\emph{microservice run-time dependency}} is the dependency from the caller microservice to the callee microservice when running.
The microservice-level dependencies complement the service-level dependencies so that the granularity of dependency management can be refined.

\section{Dependency Management System}
\label{sec:IDM}

This section elaborates on the architecture of \name demonstrated in Figure~\ref{fig:arch}.
We will introduce the data sources, the dependency analysis, and the application scenarios of \name.

\subsection{Data Source and Dependency Analysis}
\label{sec:IDM:acquisition}

The dependency information is collected from different sources.
Distributed tracing helps automatically acquire the service run-time dependency, microservice run-time dependency, and operational dependency.
By parsing the configuration files and querying the service orchestrator, \name can obtain the composed-of dependency and run-on dependency.
The engineers must report the deployment dependency of the cloud services within their duties.
Lastly, \name fuses all the dependencies for subsequent applications.

Specifically, for the microservice run-time dependency, the \name system further analyses the intensity of dependency.
Our previous work, \aid~\cite{aid}, achieves the analysis of intensity.
Specifically, given the run-time traces, \aid represents the status of each cloud service with a multivariate time series aggregated from the traces.
\aid calculates the similarities between the statuses of the caller and the callee microservices.
Finally, \aid aggregates the similarities to produce a unified value as the intensity of the dependency.
The reliability engineers will categorize the intensity into different levels by referring to the output of \aid and their domain knowledge.

\subsection{Application Scenarios}
\label{sec:IDM:applications}

In \company, \name serves hundreds of cloud services.
\name provides the engineers with full-lifecycle service reliability assistance based on the refined dependency relations.

\subsubsection{Initial Service Deployment}

According to the configured service type and dependency type, etc., \name can automatically discriminate between the compulsory and optional cloud services.
Engineers can utilize such information to assure the correct deployment of the new service.

\subsubsection{Service Upgrade}

During the service upgrade, the system is more vulnerable to new errors introduced by new versions.
Hence, avoidance of multi-point failure becomes crucial.
Before upgrading a microservice, the \name system will check the status of the cloud services and microservices it depends on.
This application scenario helps avert multi-point failures affecting changes.

\subsubsection{Architectural Optimization}

Service failures are inevitable, but \name can prevent the failures from affecting other services by optimizing improper dependencies.
\name assists in the discovery of unnecessary strong dependency on key cloud services.
If a critical service or microservice depends on another service with high intensity, \name will remind the engineers to check the necessity of the dependency.
If dependencies are not required, the development team must reduce the intensity of dependency to improve the robustness of crucial cloud services.
Since the deployment of \name, more than ten unnecessary dependencies have been discovered by \name and optimized by the development team.

\subsubsection{Failure Mitigation}

\name also assists in the mitigation of cascading failures.
During a cascading failure, \name can provide the latest intensity of dependency to On-Call Engineers (OCEs) so as to diagnose service failures efficiently.
In addition, when a cascading failure occurs, OCEs can limit the traffic to critical cloud services and recover the dependencies marked as ``strong'' first.
By doing so, the service disruption can get under control.
Once a critical failure occurs, the dependencies marked as  ``strong'' will be treated with high priority.
The failure mitigation records show that \name has reduced the time for system recovery by over 60\%.



\bibliographystyle{IEEEtran}
\bibliography{main}

\end{document}